\documentstyle[11pt,paspconf]{article}   
   
\begin{document}   
   
\title{   
H$_2$ Emission from the Inner 400 Parsecs of the Galaxy \\   
II. The UV--Excited H$_2$   
}   
   
\author{Soojong Pak\altaffilmark{1}, D. T. Jaffe,    
    and L. D. Keller\altaffilmark{1}}   
\affil{Astronomy Department, University of Texas, Austin, TX 78712}   
   
\altaffiltext{1}{   
  Visiting Astronomer, Cerro Tololo Inter-American    
  Observatory, National Optical Astronomy Observatory, which are operated    
  by the Association of Universities for Research in Astronomy,    
  under contract with the National Science Foundation}    
   
\begin{abstract}   
We have observed near--IR H$_2$ line emission on large scales in the Galactic center.
Paper~I discussed our 400~pc long strip map and 50~pc map of the H$_2$ $v=1\rightarrow 0\ S(1)$ line.
In this paper, we present observations of the higher vibrational lines (H$_2$ $v=2\rightarrow 1\ S(1)$ and $v=3\rightarrow 2\ S(3)$) at selected positions and conclude that strong far--UV radiations excites the H$_2$.
We compare the H$_2$ $v=1\rightarrow 0\ S(1)$ emission to far--IR continuum emission and show that the ratio of these two quantities in the Galactic center equals the ratio seen in the starburst galaxies, M82 and NGC~253, and in ultraluminous infrared bright galaxies. 
\end{abstract}   
   
   
   
\section{Introduction}   
   
The central kpc regions in starburst galaxies and ultraluminous IR bright    
galaxies are powerful emitters of near-IR H$_2$ emission (Puxley, Hawarden,   
\& Mountain 1990; Goldader et al. 1995).   
Ro--vibrational lines of H$_2$ can trace both photon--dominated regions (PDRs), where far--UV photons excite the H$_2$, and shocked regions, where    
the H$_2$ is thermally excited.   
Vigorous star formation in these galaxies produces large numbers of  
UV photons which fluorescently excite H$_2$, while subsequent    
supernovae shock--excite the H$_2$.   
   
We have used the University of Texas near--IR Fabry--Perot    
Spectrometer, to survey giant molecular clouds   
(GMCs) on $1-10$~pc scales 
(Luhman et at. 1994; Luhman \& Jaffe 1996; Luhman et al. 1996).   
In Orion~A, for example, the H$_2$ $v=1\rightarrow 0\ S(1)$ line emission 
extends up to 8~pc (1\deg) from the central UV source, $\theta^1$~Ori~C.   
The detection of higher vibrational state H$_2$ lines, e.g.,    
$v=6 \rightarrow 4\ Q(1)$ and $v=2 \rightarrow 1\ S(1)$, showed that far--UV photons excite the H$_2$.   
Although the shock--excited H$_2$ emission is intense in 
the Orion $BN-KL$ region,    
the emission region is relatively compact ($\sim 1\arcmin$).  
The total H$_2$ luminosity in the $BN-KL$ region is only 
$\sim 1 \%$ of the Orion PDR H$_2$ luminosity.
Similarly, UV--excited H$_2$ dominates the large--scale H$_2$ emission from other GMCs.
   
We have observed the H$_2$ emission in the inner $\sim 400$ pc ($\sim 3\deg$) of our Galaxy in order to investigate H$_2$ emission on a more global scale and to compare the Galactic center with central $\sim 1$ kpc    
regions in external galaxies.
The physical conditions in the interstellar medium of the Galactic center are    
significantly different from those in the solar neighborhood.    
The thin disk (diameter of 450~pc, height of 40~pc) of dense interstellar    
material in the Galactic center contains 
M(H$_2$) $>$ $2 \times 10^7 M_{\sun}$ (G\"{u}sten 1989; Hasegawa et al. 1996).   
The molecular clouds in the Galactic center have higher density,    
higher metallicity, and higher internal velocity dispersion than    
the clouds in the solar neighborhood (Blitz et al. 1993).   
There is strong radio continuum radiation from giant H~{\small II} regions (Sgr~A, Sgr~B, Sgr~C, and Sgr~D) and    
extended low--density (ELD) ionized gas.
The spectral index in the areas away from the discrete H~{\small II} regions shows that thermal bremsstrahlung from ionized gas can account for about half of the emission from the extended gas (Sofue 1985).   
Another indicator of the intense UV radiation in the central 400~pc is strong far--IR continuum emission (Odenwald \& Fazio 1984).
About 90\% of the far--UV energy    
is absorbed by dust and reradiated in the far--IR.   
From the far--IR intensity, we estimate that the far--UV radiation field is  
$\sim 10^3$ times the value in the solar neighborhood    
($I_\circ = 4 \times 10^{-4}$ ergs~s$^{-1}$~cm~$^{-2}$~sr$^{-1}$,    
Draine 1978).   
The energetic conditions in the Galactic center mean that the center can   
provide a unique view of the interaction between stellar UV radiation and    
molecular clouds, and serve as a nearby model for the nuclei of    
galaxies.   

\begin{figure}[t]
\vbox to 7.9cm {
  \plotfiddle{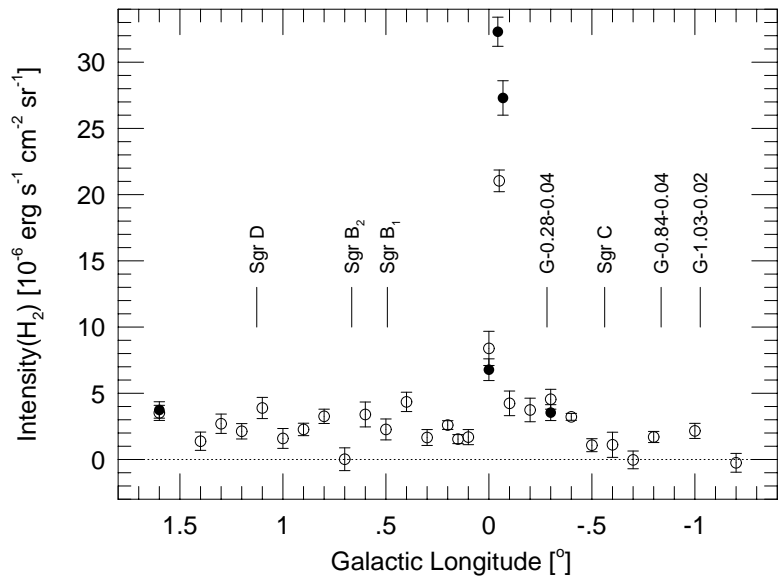} {7.9cm} {0} {40} {40} {-160} {0}
}
\caption{ \label{fig-1}
Observed intensity distribution of H$_2$ $v=1\rightarrow 0\ S(1)$
($\lambda = 2.121\ \micron$)
along the Galactic plane at $b = -0\fdg05$.
The open circles were taken at the McDonald 0.9 m telescope with a 3\farcm3
beam (Paper I) and the filled circles at the CTIO 1.5 m telescope with a
1\farcm35 beam.
The intensities have not been corrected for interstellar extinction.
The error bars represent $1\sigma$ measurement uncertainties.
}
\end{figure}
   
In paper I (Pak, Jaffe, \& Keller 1996) we showed the distribution of
H$_2$ $v=1\rightarrow 0\ S(1)$ emission along a 400 pc--long    
strip and in the inner 50~pc of the Galactic center.   
We detected H$_2$ emission throughout the surveyed region.    
The typical dereddened ($A_K = 2.5$ mag) H$_2$ $v=1\rightarrow 0\ S(1)$    
intensity, 
$\sim 3 \times 10^{-5}$ ergs~s$^{-1}$~sr$^{-1}$,   
is similar to the surface brightness in Galactic PDRs (Luhman \& Jaffe 1996).    
In this Paper, we present observations of several H$_2$ lines, discuss the    
excitation mechanism, and compare the Galactic center observations to observations of other galaxies.   
   
\section{Observations and Results}   
  
We observed three H$_2$ emission lines:    
$v=1\rightarrow 0\ S(1)$ ($\lambda = 2.121\ \micron$),   
$v=2\rightarrow 1\ S(1)$ ($\lambda = 2.247\ \micron$), and   
$v=3\rightarrow 2\ S(3)$ ($\lambda = 2.201\ \micron$),   
at the Cerro Tololo Inter--American Observatory 1.5 m telescope in 1995 July and October.   
We used the University of Texas Near--Infrared Fabry--Perot Spectrometer.
The instrument was specially designed to observe very extended, low surface 
brightness objects, and has a single channel InSb detector with surface area 
of 1~mm to maximize the beam size (Luhman et al. 1995).   
The telescope ($f/30$), a collimator (effective focal length 686 mm), and    
a field lens (effective focal length 20mm) produce a beam diameter    
of 1\farcm 35 (equivalent disk).   
   
The Fabry--Perot interferometer operates in 94th order    
($\lambda_\circ = 2.121 \micron$) with an effective finesse of 26, yielding a    
spectral resolution of 125 km~s$^{-1}$ (FWHM).   
Scans covered in 15 sequential steps, $\pm 300$~km~s$^{-1}$ centered at    
$V_{LSR} \simeq 0$~km~s$^{-1}$.
In order to subtract background and telluric OH line emission, we chopped    
the secondary mirror to $\Delta b$ = $+16\arcmin$ or $-16\arcmin$ at 0.5~Hz.

We observed five positions: ($l$, $b$) = 
($-0\fdg0433$, $-0\fdg0462$), 
($-0\fdg0683$, $-0\fdg0462$),   
($0\fdg00$, $-0\fdg05$), 
($-0\fdg30$, $-0\fdg05$), and 
($+1\fdg60$, $-0\fdg05$).
In Figure~\ref{fig-1}, we plot the new H$_2$ $v=1\rightarrow 0\ S(1)$ data 
overlaid on the data from Paper I and compare the two data sets.   
The 3\farcm3 beam of the McDonald 0.9 m telescope centered at Sgr A$^*$    
($l=-0\fdg0558,\ b=-0\fdg0462$) covers the whole circumnuclear gas ring (Gatley et al. 1986), while, with the 1\farcm35 beam of the CTIO 1.5 m telescope, we observed 
the $+\Delta l$ H$_2$ peak ($-0\fdg0433,\ -0\fdg0462$) and    
the $-\Delta l$ H$_2$ peak ($-0\fdg0683,\ -0\fdg0462$).
The difference between the 3\farcm3 beam data and the 1\farcm35 beam data toward Sgr~A is an effect of different beam sizes because the H$_2$ emission sources are relatively compact.
In the large--scale emission beyond Sgr~A, the two data sets agree to within the errors, indicating that the H$_2$ emission varies slowly on $1\arcmin - 3\arcmin$ scales.
   
\section{ Extinction Correction } \label{sec:extinction}   
   
\begin{figure}[p]   
\vbox to 8cm {   
  \plotfiddle{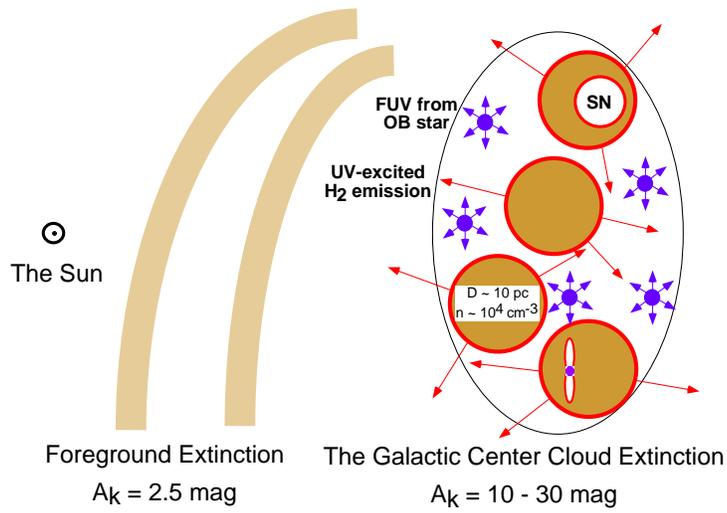} {8cm} {90} {40} {40} {140} {20}   
}   
\vbox to 8cm {   
  \plotfiddle{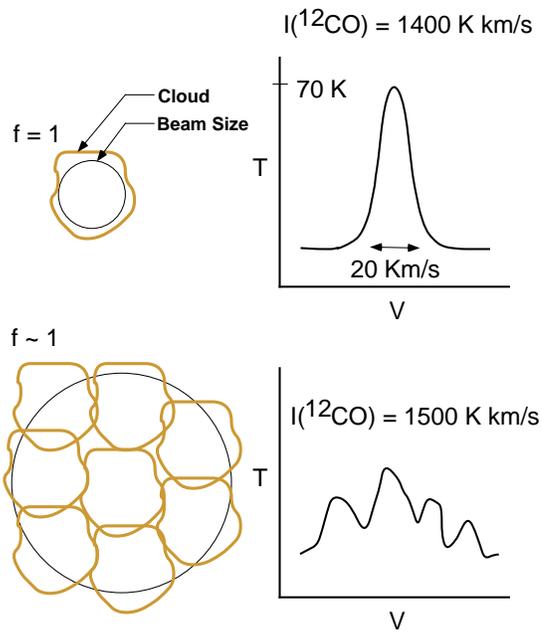} {8cm} {0} {40} {40} {-120} {-28}   
}   
\caption{ \label{fig-2}   
 (a) Top--view schematic of the distribution of interstellar material in two foreground spiral arms (foreground extinction) and in the GMCs in the inner $\sim 400$~pc of the Galaxy (Galactic center extinction).
 (b) Schematic diagram of small and large beam observations in the Galactic center. 
$^{12}$CO $J = 1\rightarrow 0$ spectrum of a typical cloud is beam diluted.
The velocity--integrated intensity including the clouds in the beam at other velocities is $\sim 1500$ K~km~s$^{-1}$,
which indicates that the area filling factor, $f$, is $\sim 1$.
The cloud components do not usually overlap along the line--of--sight.
}   
\end{figure}   

\begin{figure}[t]
\vbox to 8.2cm {
  \plotfiddle{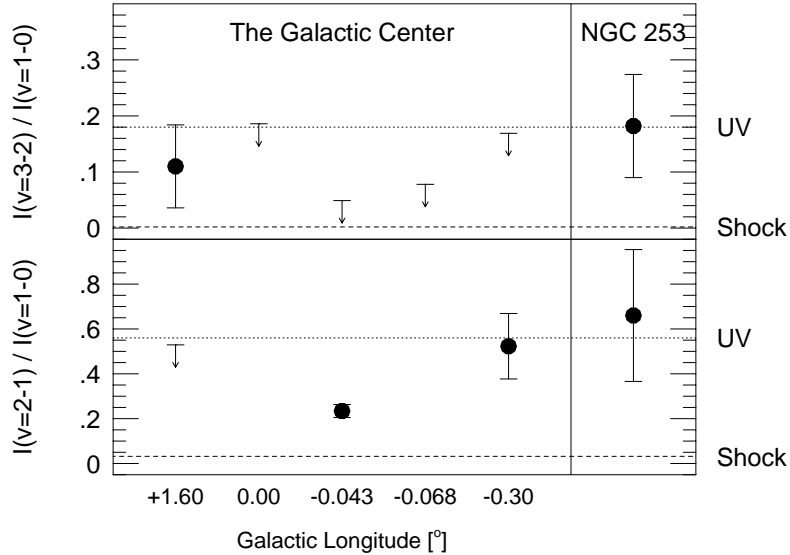} {8.2cm} {0} {40} {40} {-160} {0}
}
\caption{ \label{fig-3}
Observed H$_2$ line ratios at positions along the Galactic Plane
($b=-0\fdg05$) and in the central 1 kpc of NGC 253.
The dotted lines are modeled ratios of UV--excited H$_2$ lines,
(Black \& van Dishoeck 1987), and
shock--excited H$_2$ lines
($V_{shock} = 30$ km~s$^{-1}$; Draine, Roberge, \& Dalgarno 1983).
The arrows show the $3\sigma$ limits where we did not detect the higher
vibrational level lines.
}
\end{figure}

\begin{figure}[t]
\vbox to 8cm {
  \plotfiddle{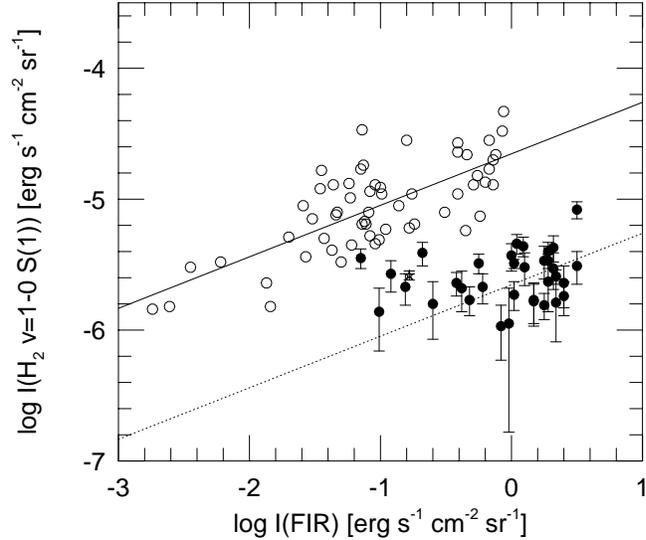} {8cm} {0} {40} {40} {-155} {0}
}
\caption{ \label{fig-4}
$I_{FIR}$ versus $I_{H2 v=1\rightarrow 0\ S(1)}$ for the Galactic PDRs
and the Galactic center.
The open circles are from Orion~A and B, $\rho$~Ophiuchi, and G236+39
(Luhman \& Jaffe 1996), and
the filled circles are from the Galactic center (Paper I).
The Galactic center data are not corrected for extinction.
The solid line
(log$I_{H2\ v=1\rightarrow 0\ S(1)}$ = $-4.65$ + 0.39~log$I_{FIR}$)
is derived from the Galactic PDR data using a least squares method, and
the dotted line shows the vertically shifted solid line by
$\Delta$log$I_{H2}$ = $-1$.
}
\end{figure}

\begin{figure}[t]
\vbox to 9cm {
  \plotfiddle{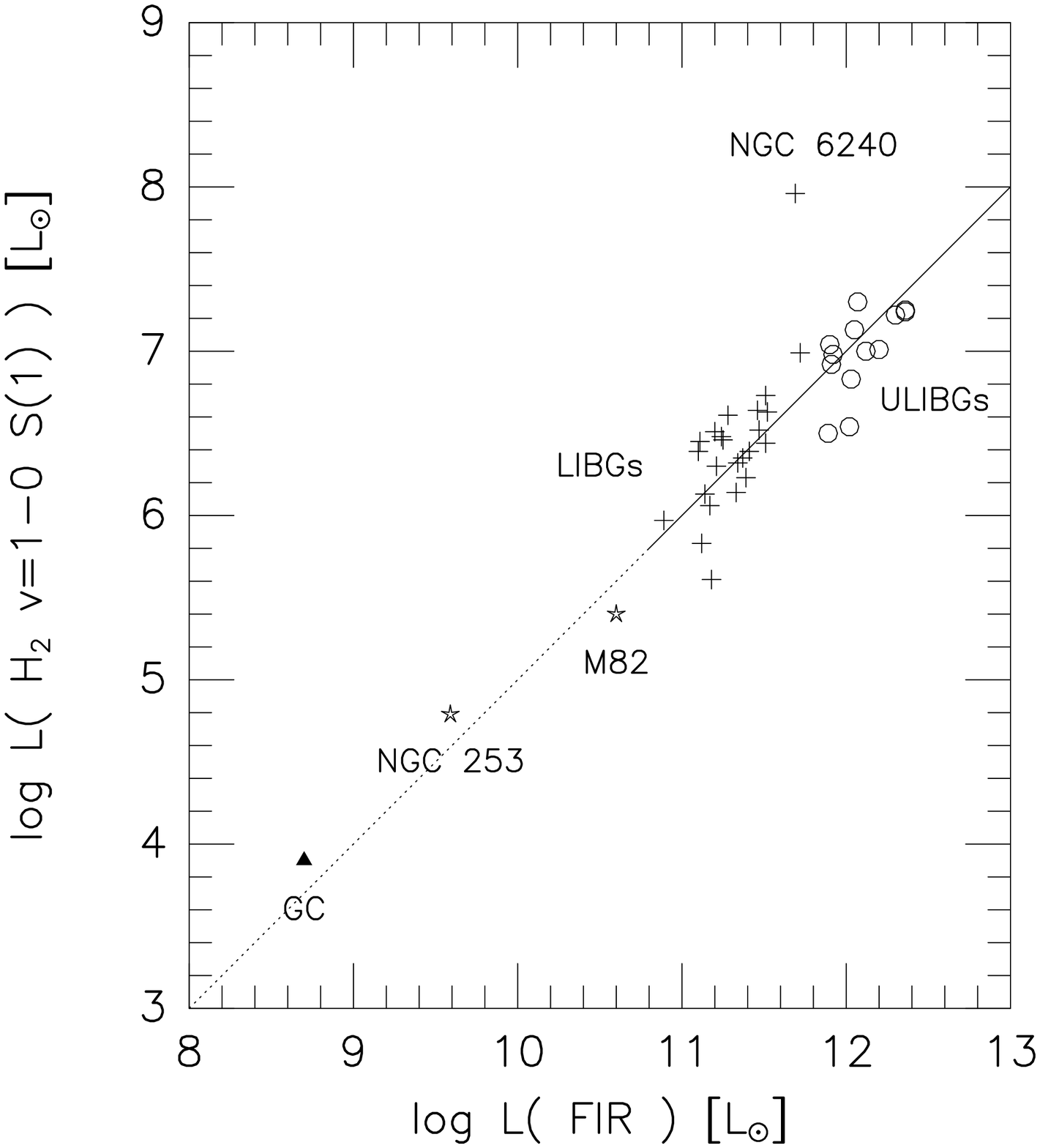} {9cm} {0} {40} {40} {-125} {-20}
}
\caption{ \label{fig-5}
$L_{FIR}$ versus $L_{H2\ v=1\rightarrow 0\ S(1)}$ of various kinds of
galaxies.
The solid line
(log$L_{H2\ v=1\rightarrow 0\ S(1)}$ = $-5$ + log$L_{FIR}$)
is derived from data of
ultraluminous IR bright galaxies (open circles) and
luminous IR bright galaxies (plus signs, Goldader et al. 1995)
The dotted line shows extrapolation from the solid line.
The H$_2$ data of M82 were taken at the McDonald 2.7 m telescope
and the H$_2$ data of NGC 253 at the CTIO 1.5 m telescope, both with the UT~FPS.
}
\end{figure}
  
The extinction in K--band toward the Galactic center is significant.   
Figure~\ref{fig-2}a shows the classification of the extinction into 
{\it foreground extinction} by material in spiral arms at $R=4-8$ kpc, and    
{\it Galactic center extinction} by material in the Galactic center clouds.   
Catchpole, Whitelock, \& Glass (1990) measured the foreground extinction as    
$A_K \simeq 2.5$~mag.   
   
A discussion of the Galactic center extinction requires a different approach
because individual clouds in the Galactic center are almost    
opaque in the near--IR ( $A_K = 10-30$ mag for typical clouds of
$D$ $\simeq$ 10~pc and $n$($H_2$) $\simeq$ $10^4$~cm$^{-3}$).   
If the UV--excited H$_2$ emission arises on the cloud    
surfaces, we need only consider the effects of shadowing by other Galactic center clouds (see Figure~\ref{fig-2}b).   
From millimeter observations of $^{12}$CO $J = 1\rightarrow 0$ emission,
we can estimate the velocity--integrated area filling factor of clouds, $f$.
If the millimeter telescope beam size is smaller than the individual clouds and covers only one cloud along the line--of--sight, the area filling factor, $f$, is 1.
The upper diagram in Figure~\ref{fig-2}b shows
an expected $^{12}$CO $J = 1\rightarrow 0$ spectrum of
typical clouds in the Galactic center which have 
kinetic temperature of $\sim 70$ K and line widths of $\sim 20$ km~s$^{-1}$
(G\"{u}sten 1989).
In general, the clouds have different sizes and may overlap along 
the line--of--sight.
The lower diagram in Figure~\ref{fig-2}b shows an observed typical 
$^{12}$CO $J = 1\rightarrow 0$ spectrum where the
velocity--integrated intensity is $\sim 1500$ K~km~s$^{-1}$ 
(Bally et al. 1987; Bally et al. 1988).
The value $f$ is the ratio of the observed velocity--integrated intensity of
$^{12}$CO $J=1\rightarrow 0$ to the single typical cloud intensity
(70~K $\times$ 20~km~s$^{-1}$).
The $f$ toward the Galactic center clouds is $\sim 1$,
implying that there is little or no overlap along a typical line--of--sight. 
If $f \leq 1$, we only miss the near--IR H$_2$ flux from the back sides of the clouds.
If $f > 1$, H$_2$ radiation is blocked by the foreground clouds, and the ratio of the observed H$_2$ flux to the emitted flux is inversely proportional to $f$.
Since $f \simeq 1$, we use the foreground values, $A_K = 2.5$, for the extinction correction.   

\section{ H$_2$ Excitation Mechanism }   
   
\subsection{ H$_2$ Line Ratios }   
   
In UV--excited H$_2$, the branching ratios in the downward cascade determine the relative strengths of the near--IR lines.
On the other hand, 
the energy level populations of shock--excited H$_2$ are thermalized.   
We use the line intensity ratios of higher vibrational level lines to    
the $v=1\rightarrow 0\ S(1)$ line in order to identify the H$_2$ excitation 
mechanism.
   
In Figure~\ref{fig-3}, the observed ratios in the large--scale Galactic center  
and the central 1 kpc region of NGC~253 imply that the H$_2$ emission may  
result from UV--excitation.  
In the circumnuclear gas ring ($l=- 0\fdg0433$ and $-0\fdg0683$),    
the UV--excited H$_2$ energy levels are partially    
thermalized because of the relatively high density   
(Sternberg \& Dalgarno 1989; 
see also Ramsay-Howat, Mountain, \& Geballe 1996 for the H$_2$ observations in the circumnuclear gas ring).   
The determination of line ratios consistent with UV excitation in the large--scale Galactic center and NGC~253 means the gas is not dense enough for collisions to significantly alter the radiative cascade, 
n(H$_2$) $<$ $10^5$~cm$^{-3}$ (Luhman et al. 1996).

\subsection{ $I_{FIR}$ versus $I_{H2}$ }  
  
If large--scale H$_2$ emission arises in the surface layers of the clouds  
where far--UV photons can excite the molecules, the dust, which absorbs the  
bulk of the incident flux, ought to radiate in the far--IR continuum as well.  
If we de--redden the Galactic center H$_2$ observations  
by $A_K$ = 2.5~mag, the Galactic center results are consistent  
with the empirical far--IR vs. H$_2$ relationship derived for the   
UV--excited surfaces of clouds in the galactic disk (see Figure~\ref{fig-4}).   
  
\section{ Comparison with other Galaxies }   
  
We extrapolate from our 400 pc long strip to the total   
H$_2$ $v=1\rightarrow 0\ S(1)$ luminosity of the Galactic Center by assuming that the scale height of the H$_2$ emission equals that of the far--IR radiation ($h\simeq 0\fdg2$, Odenwald \& Fazio 1984)   
and that $A_K = 2.5$~mag and $f \simeq 1$.  
The H$_2$ $v=1\rightarrow 0\ S(1)$ luminosity in the inner 400~pc diameter   
of the Galaxy is $8.0 \times 10^3 \ L_{\sun}$.   
  
For ultraluminous and luminous  
infrared bright galaxies ($L_{IR}$ $\ga$ $10^{11}\ L_{\sun}$),  
Goldader et al. (1995) showed the correlation between $L_{FIR}$ and  
$L_{H2\ v=1\rightarrow 0\ S(1)}$.  
We can extend the relationship to nearby starburst galaxies like M82 and   
NGC 253, and to the Galactic center (see Figure~\ref{fig-5}).  
The strong correlation between the far-IR and H$_2$ luminosity for various   
classes of galaxies indicates that the far--UV radiation may excite large   
scale H$_2$ emission in all of these sources.

\acknowledgments   
   
This work was supported by NSF grant AST 9117373 and by David and Lucile Packard Foundation.  
We thank M. Luhman and T. Benedict for contributions to the Fabry--Perot   
Spectrometer Project, and J. Elias, B. Gregory, and the staff of the CTIO for   
their assistance in setting up our instrument.

\end{document}